\pdfoutput = 1
\documentclass{PoS}
\newcommand{\KK}{KK}
\title{KKMC-hh: A Precision Event Generator for EW Radiative Corrections in Hadron Scattering}
\ShortTitle{KKMC-hh}
\author{\speaker{S.A. Yost}\thanks{This presentation was supported by a grant from the Citadel Foundation.}\\
        Author The Citadel, The Military College of South Carolina\\
        E-mail: \email{scott.yost@citadel.edu}}
\author{S. Jadach\\
        Institute of Nuclear Physics IFJ-PAN, Cracow, Poland\\
        E-mail: \email{stanislaw.jadach@cern.ch}}
\author{B.F.L. Ward\\
	Baylor University\\
	E-mail: \email{bfl\_ward@baylor.edu}}
\author{Z. W\c{a}s\\
        Institute of Nuclear Physics IFJ-PAN, Cracow, Poland\\
        Email: \email{wasm@cern.ch}}
\abstract{KKMC-hh is a precision event-generator for $Z$ production and decay in
hadronic collisions, which applies amplitude-level exponentiation to both 
initial and final state photon radiation, including perturbative residuals 
through order $\alpha^2 L$, together with electroweak matrix element 
corrections. We present results showing 
the effect of multi-photon radiation for cuts motivated by a recent ATLAS 
$W$-mass analysis. We also show preliminary untuned comparisons of the 
electroweak corrections of KKMC-hh to those of HORACE, which includes order
$\alpha$ corrections with exponentiated final-state photon radiation.
\\[1in]
\centerline {\bf BU-HEPP-18-01, January 2018}
}
\FullConference{13th International Symposium on Radiative Corrections\\
		24-29 September, 2017\\
		St. Gilgen, Austria}
\begin{document}
\section{Introduction}
A precise  calculation of vector boson production is essential for interpreting
electroweak precision data from the LHC and anticipated future hadron colliders.
For example, a recent analysis~\cite{atlasmw-17} of the 7 TeV data from ATLAS 
yielded a measure
of the $W$ boson mass of 
$$M_W = 80370\pm 7 (\rm{stat.}) \pm 11 (\rm{exp. syst.}) \pm 14 (\rm{mod. syst.})\ \rm{MeV}\\
=   80370 \pm 19\ \rm{MeV}.$$
The modeling systematic error of 14 MeV is the largest contribution to the 
uncertainty.
 With increasing statistics, it will become important to reduce
this modeling error. Certain aspects of the $W$ production and decay 
systematics were estimated by comparing to analogous systematics for 
$Z/\gamma^*$ production and decay, so improving the electroweak precision 
for the $Z$ calculation improves the $W$ mass result as well.

ATLAS has reported~\cite{atlas-1612-03016} 
a measurement of the $Z$ differential spectra with a 2.5 per
mille statistical error in the peak region, and comparable NLO EW corrections
at central rapidities when FSR is unfolded using 
PHOTOS.~\cite{Barberio:1990ms,Barberio:1994qi,Golonka:2005pn} 
Per-mille level statistical errors in differential spectra for $Z/\gamma^*$ 
production and decay are also reported by ATLAS in Ref.\ \cite{atlas-8TeV}, 
again unfolding FSR effects using PHOTOS. The analysis of 
Ref.\ \cite{atlas-7TEV} shows a 0.2\% statistical
error in the differential spectra for $Z/\gamma^*$ production for 7 TeV data at
the LHC. Unfolding the FSR with PHOTOS (cross-checked with SHERPA~\cite{sherpa})
results in a
0.3\% error in the $P_{\rm T}$ spectrum for muon pairs, and a 0.1\% error for
electron pairs. CMS has also reported per mille level statistical errors in the
$Z/\gamma^*$ $P_{\rm T}$ spectrum, with a per mille level estimate of the 
systematic error due to FSR.~\cite{cms-8TEV} 

These increasingly precise analyses 
raise the question of what per-mille level higher-order EW corrections should be
included in a careful analysis of the systematics.  With 
\KK{MC}-hh,~\cite{kkmchh,herwiri2} it is possible to answer this question by 
unfolding ${\cal O}(\alpha^2 L)$ EW corrections from the data. This would give 
a more complete test of the contributions relevant to precision EW measurements.
We will present results from \KK{MC}-hh showing the result of calculating 
multi-photon effects at various levels of precision for cuts inspired by the
ATLAS $W$-mass analysis, as well as comparisons to other programs, including
HORACE~\cite{Horace1,Horace2}, which calculates NLO EW corrections with 
exponentiated FSR. 

\section{The Physics of \KK{MC}-hh}
\KK{MC}-hh~\cite{kkmchh,herwiri2} is based on the LEP-era event generator 
\KK{MC}~\cite{Jadach:1999vf} for 
$e^+ e^-\rightarrow f{\overline f} + n\gamma$, where $f{\overline f}$ represents
a final state fermion pair, for CMS energies from $2m_\tau$ to 1 TeV. The
precision tag for LEP2 was 0.2\%. The MC structure is based on 
CEEX,~\cite{Jadach:1993yv,Jadach:1999vf,Jadach:2000ir,Jadach:2013aha} an 
amplitude-level analog of YFS exponentiation,~\cite{yfs} 
and includes residuals through
order $\alpha^2 L$ where $L - \ln(s/m_e^2)$ is a relevant ``big logarithm.'' 
Electroweak matrix element corrections are included via DIZET 6.21, from the
semi-analytical program ZFITTER~\cite{zfitter6:1999}. DIZET calculates 
vacuum polarization factors
for the photon and $Z$ propagators, and adds form-factor corrections to the 
vector coupling and an angle-dependent form-factor to incorporate the effect
of box diagrams. The correction factors are tabulated at the beginning of a 
run.  Tau decay is implemented using 
TAUOLA.~\cite{Jadach:1990mz,Jezabek:1991qp,Jadach:1993hs,Golonka:2003xt} 

Version 4.22~\cite{Jadach:2013aha} of \KK{MC} supports quark
initial states, and a modified version of this is incorporated into \KK{MC}-hh
with the addition of support for selecting the quarks via PDFs, using an 
LHAPDF\cite{lhapdf} interface. 
\KK{MC}-hh uses the adaptive MC program FOAM~\cite{foam} to to generate the 
quark momentum fractions $x_i$, the total ISR energy, and the quark flavor 
using a crude distribution which is constructed during an initialization phase. 

CEEX was introduced~\cite{Jadach:2000ir} to overcome limitations of traditional 
YFS exponentiation,
which suffered from a proliferation of interference terms, limiting its 
ability to reach the desired 0.2\% precision tag for LEP2. CEEX works at the
level of spinor helicity amplitudes, which greatly facilitates calculating 
effects such as ISR-FSR interference (IFI). The IFI effects in \KK{MC}-hh can 
be switched on or off, allowing an assessment of their importance. Also, it is
possible to switch to traditional YFS exponentiation, called EEX, for comparison
purposes. The level of residuals can also be selected, so that the effect of 
higher-order corrections in $\alpha$ can be assessed.

\KK{MC}-hh can export its events in an LHE~\cite{lhe-format} 
event record to be showered externally,
or the events can be showered internally using HERWIG 6.5.~\cite{HERWIG} 
In the following 
sections, we will present both showered and unshowered results.  There are
also plans to implement a mode in which \KK{MC}-hh can reweight events from an
any external QCD shower generator, and add photons. The combination of NLO
QCD with EW corrections will take advantage of the fact that to a good 
approximation, these corrections factorize.~\cite{den-ditt1211.5078,dittmr} 
The direct inclusion of NLO QCD into \KK{MC}-hh is also anticipated, using the 
KrkNLO~\cite{krknlo} scheme. 

\section{\KK{MC}-hh Results for the ATLAS Acceptance}

As noted in the introduction, the recent ATLAS measurement of $M_W$ estimated
some aspects of the $W$ production and decay systematics, such as the 
uncertainty in the momentum resolution scale, using the analogous systematics
for $Z/\gamma^*$ events. The uncertainty in the EW corrections in $Z$ production
thus contributes to the $W$ mass measurement's systematics. In this section, we
illustrate the size of the higher order EW effects now available via 
\KK{MC}-hh using the cuts applied in the ATLAS analysis. More details of this
analysis may be found in Ref.\ \cite{kkmchh-atlas}. The 
ATLAS cuts~\cite{atlasmw-17} on the 
invariant mass and transverse momentum of the dilepton pair are 
$$80 {\rm GeV} < M_{\ell\ell} < 100 {\rm GeV}, P_{\rm T}^{\ell\ell} 
< 30 {\rm GeV},$$
while the transverse momentum and pseudorapidity of each lepton is constrained 
by 
$$ P_{T}^{\ell} > 25 {\rm GeV}, |\eta_\ell| < 2.5.$$

In these tests, we generate $10^8$-event samples for 7 TeV collisions 
using MSTW2008 PDFs~\cite{mstw2008} and 
shower with HERWIG6.5.  We compare the best ${\cal O}(\alpha^2 L)$ CEEX
implementation (labeled CEEX2) to several more limited models: 
${\cal O}(\alpha^2 L)$ CEEX without IFI (initial-final interference), 
${\cal O}(\alpha)$ EEX (labeled EEX1), and 
${\cal O}(\alpha)$ EEX without ISR (initial-state radiation). Table 1
table shows the cross-sections with and without the cuts. 
All of the 
calculations are compatible to a fraction of a per mille, 
The cut cross section with ISR off shows a per mille difference relative to
the full CEEX2 result, while all other differences are a fraction of a per
mille. 

\vbox{
\begin{center}
\begin{tabular}{|l|c|c|c|c|}
\hline
	    	& uncut (pb) & Difference	& cut (pb) & Difference\\
\hline
CEEX2 	    	& 844.74     	& $\times$  & 280.36 & $\times$\\
CEEX2 (no IFI) 	& 844.97 	& $+0.03\%$ & 280.31 & $-0.02\%$\\
EEX1  		& 844.45 	& $-0.03\%$ & 280.38 & $+0.007\%$\\
EEX1 (no ISR)  	& 844.97 	& $+0.03\%$ & 280.64 & $+0.10\%$\\
\hline
\end{tabular}
\\[1em]
{{\bf Table 1.} Total Cross Sections With and Without ATLAS Cuts. Differences
are shown relative to CEEX2.}
\end{center}
}

Figures 1 and 2 compare differential spectra for the 
transverse momentum and pseudorapidity of the $\mu^-$ in $Z/\gamma^*$ production
with decay to muon pairs for the ATLAS cuts. While the contribution of ISR to
the total cross section was at the per mille level, the effect on distributions
is much greater, especially in the $P_{\rm T}$ distribution where a several 
percent ISR effect is seen. Initial-final interference (IFI) is a fractional
per mille effect. 

\begin{figure}[h!]
\begin{center}
\setlength{\unitlength}{1in}
\begin{picture}(6.5,3.0)
\put(0,0.5){\includegraphics[width=3.2in,height=2.6in]{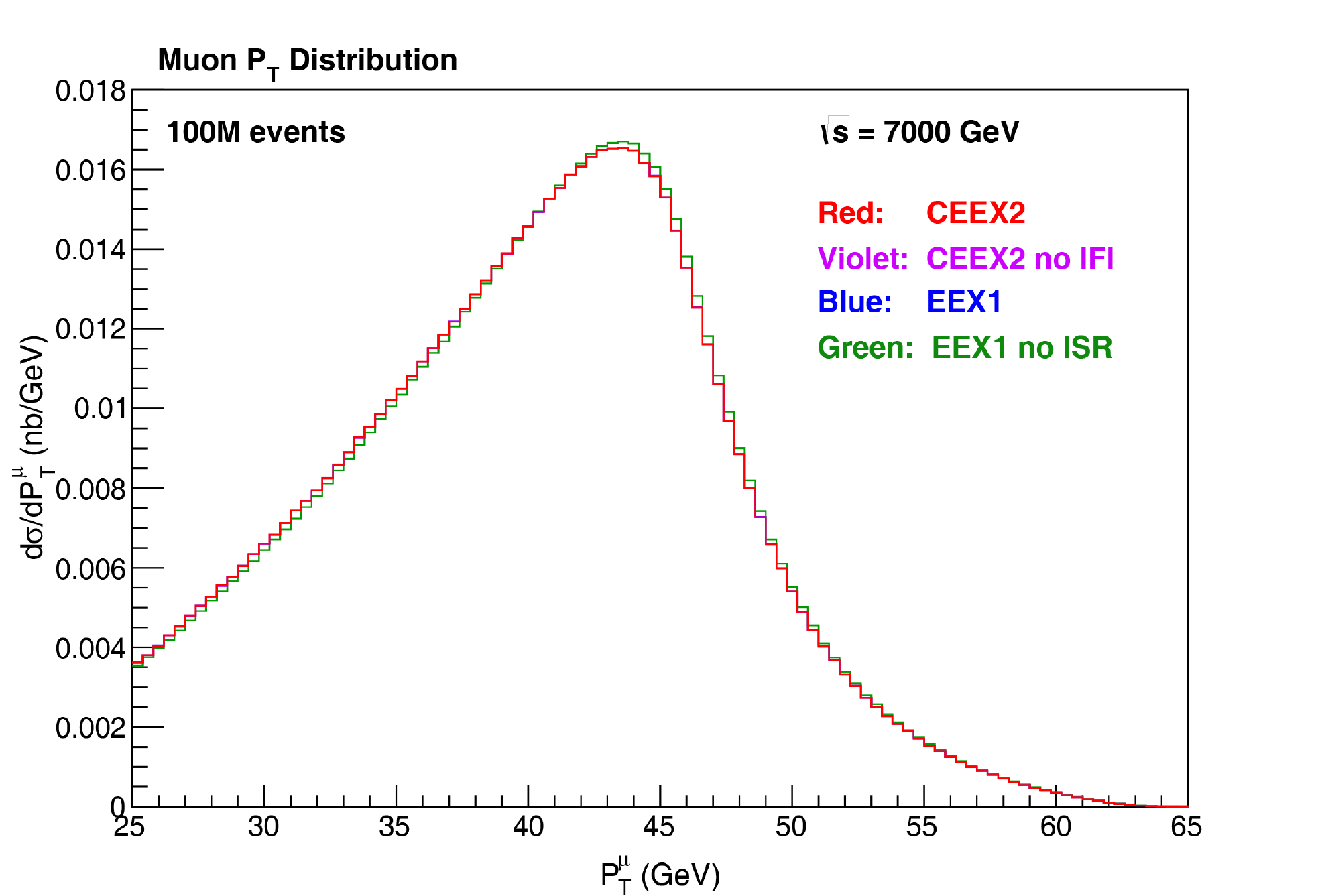}}
\put(3.2,0.5){\includegraphics[width=3.2in,height=2.6in]{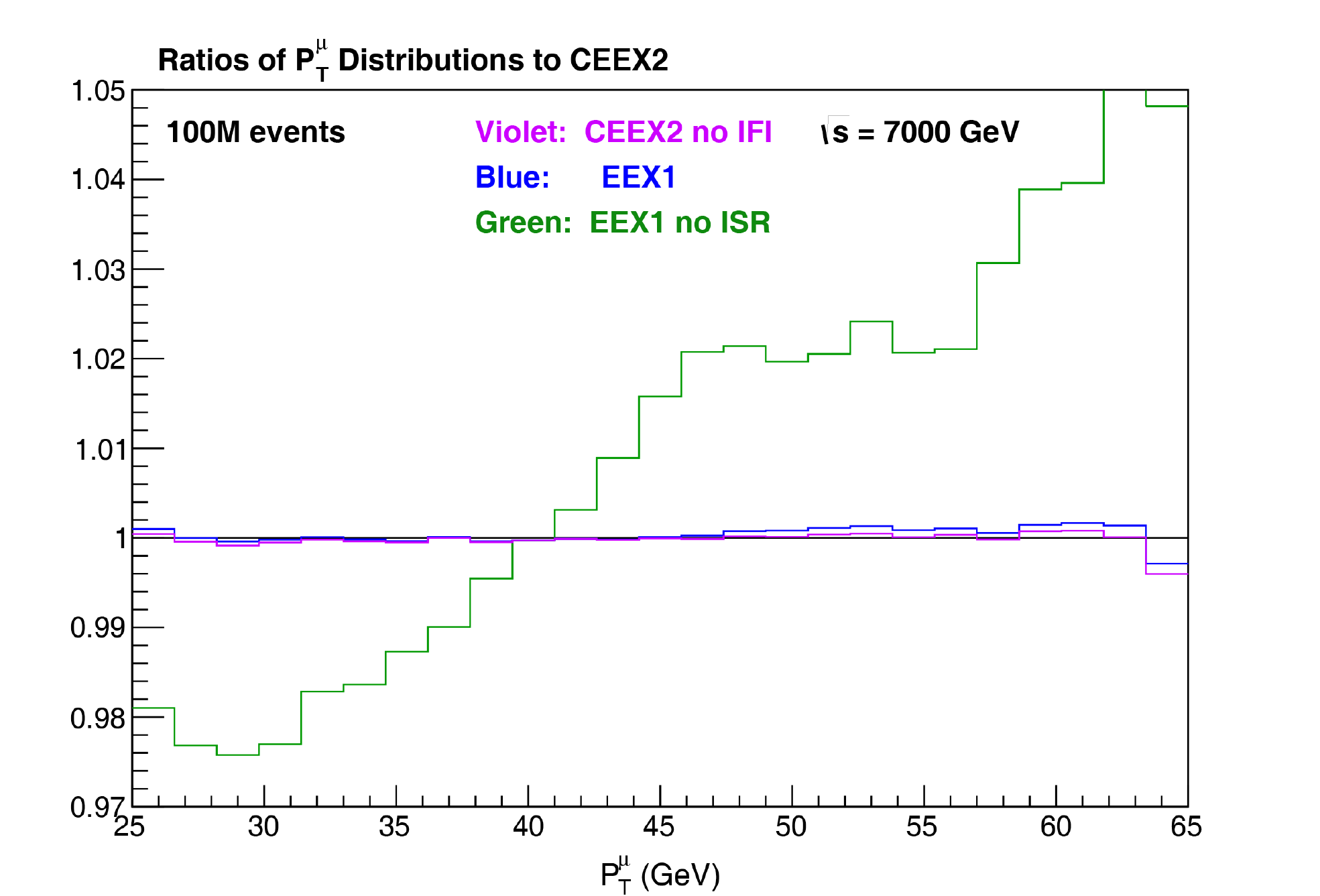}}
\end{picture}
\vspace{-0.75in}
\caption{Muon Transverse Momentum Distributions and Ratios}
\end{center}
\end{figure}

\begin{figure}[ht!]
\begin{center}
\setlength{\unitlength}{1in}
\begin{picture}(6.5,3.0)
\put(0,0.5){\includegraphics[width=3.2in,height=2.6in]{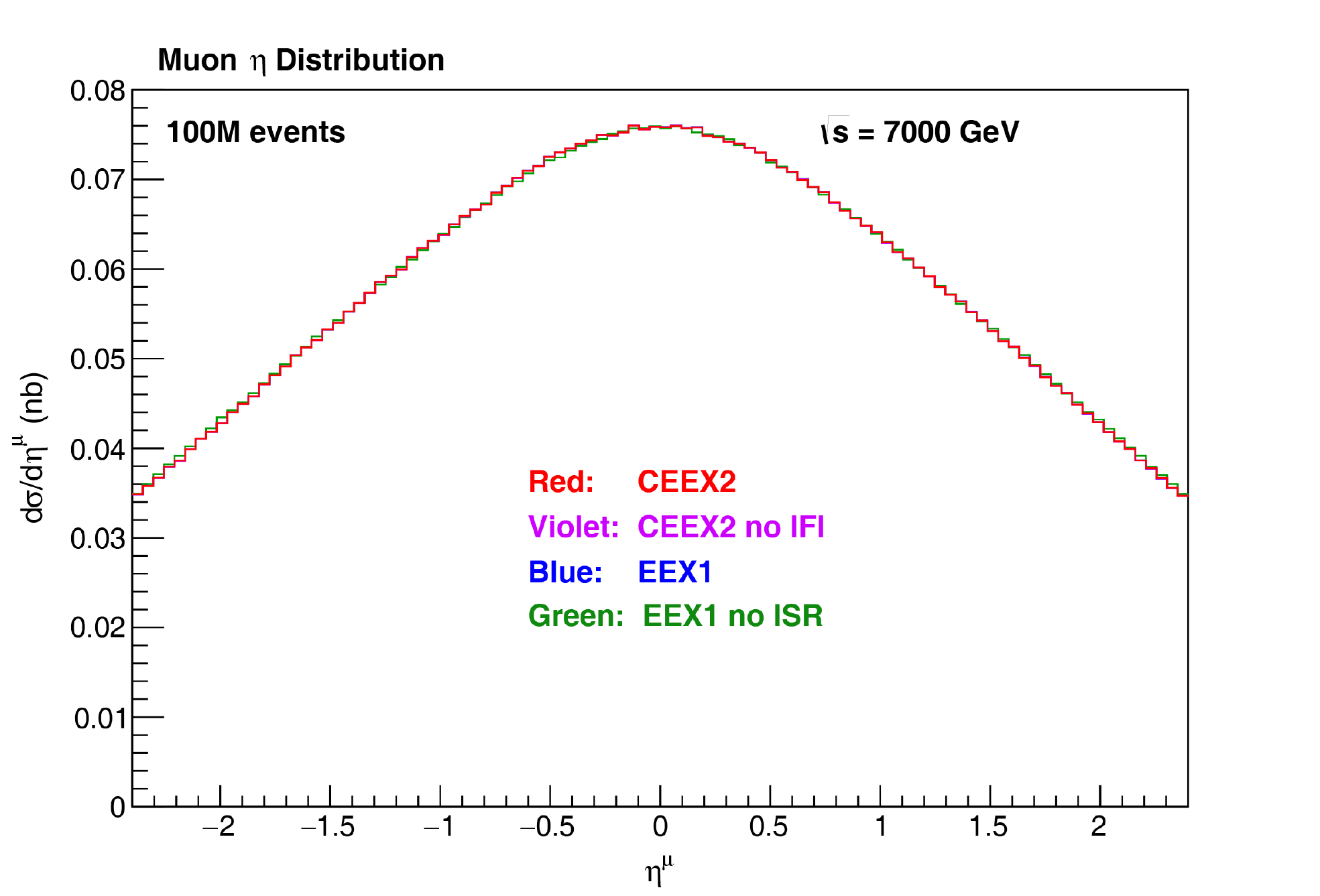}}
\put(3.2,0.5){\includegraphics[width=3.2in,height=2.6in]{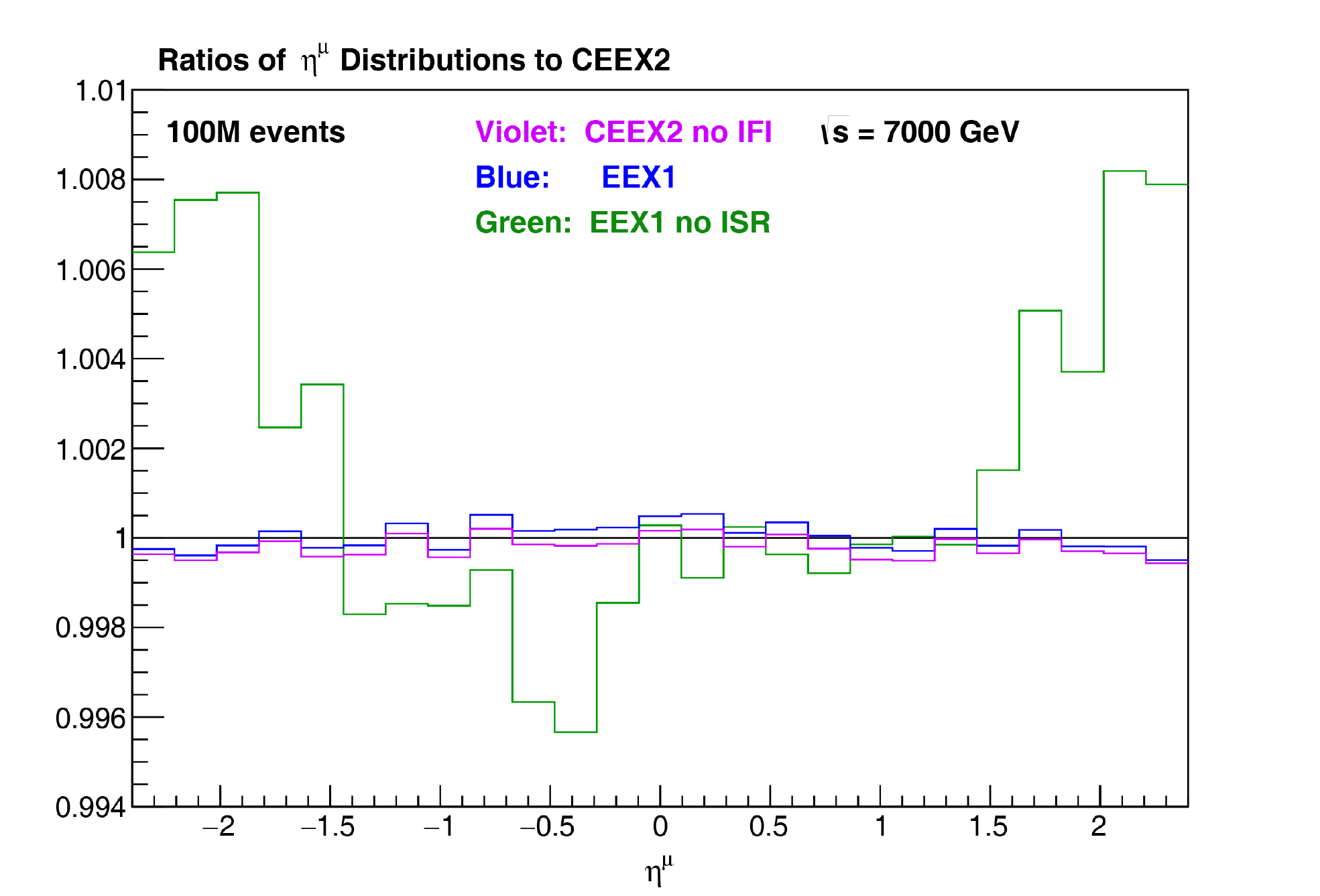}}
\end{picture}
\vspace{-0.75in}
\caption{Muon Pseudorapidity Distributions and Ratios}
\end{center}
\end{figure}

Figures 3 -- 5
compare the dimuon invariant mass, transverse momentum, and rapidity for these
cuts. Photon distributions are shown in figures 6 and 7.
Percent-level ISR effects are also seen in the 
dimuon invariant mass spectrum. Initial-final interference is a fractional per 
mille effect in all cases. In calculations were per mille level accuracy is 
required, all of the contributions in \KK{MC}-hh should be taken into account.

\begin{figure}[hb!]
\begin{center}
\setlength{\unitlength}{1in}
\begin{picture}(6.5,3.0)
\put(0,0.5){\includegraphics[width=3.2in,height=2.6in]{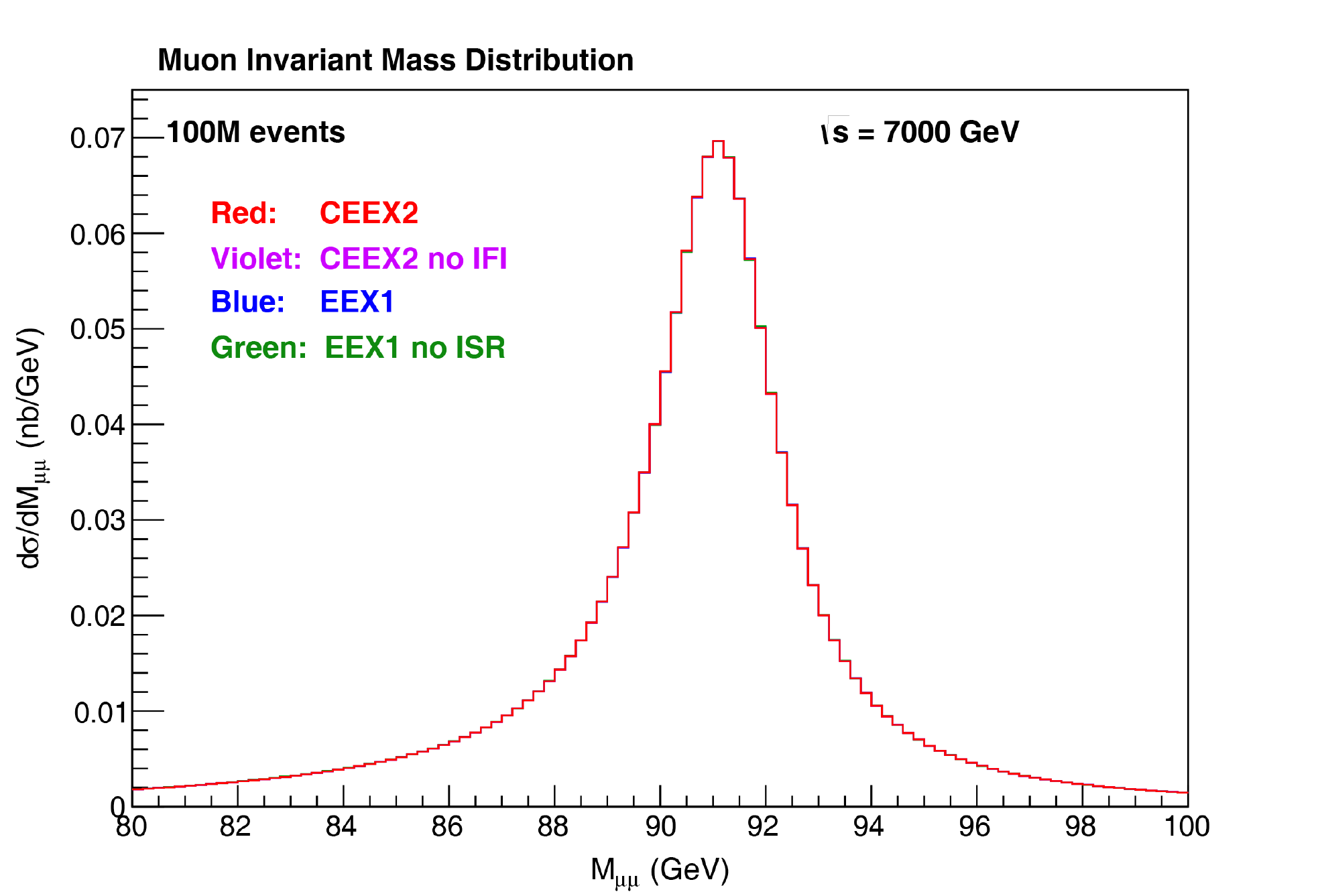}}
\put(3.2,0.5){\includegraphics[width=3.2in,height=2.6in]{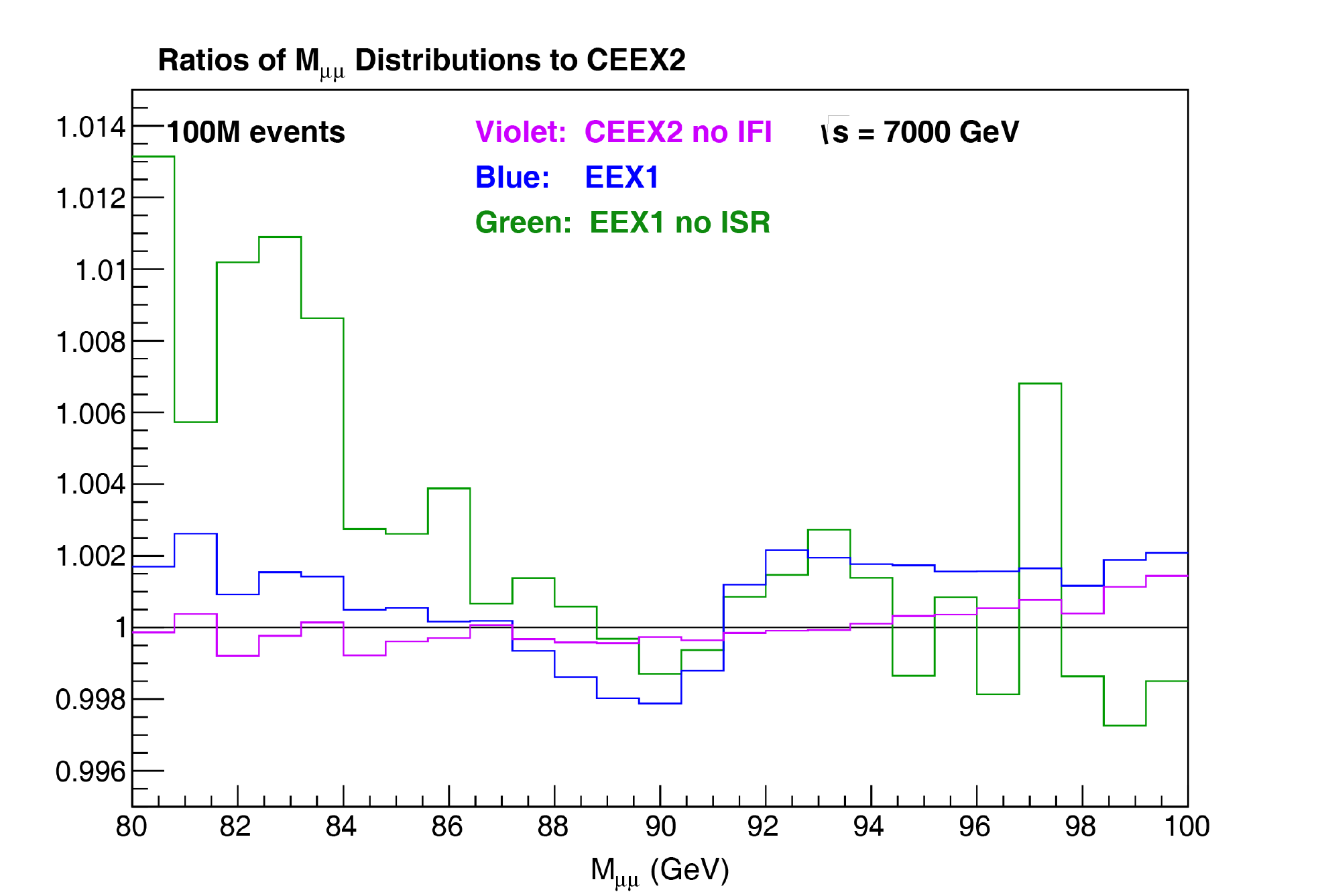}}
\end{picture}
\vspace{-0.75in}
\caption{Dimuon Invariant Mass Distributions and Ratios}
\end{center}
\end{figure}

\begin{figure}[ht!]
\begin{center}
\setlength{\unitlength}{1in}
\begin{picture}(6.5,3.0)
\put(0,0.5){\includegraphics[width=3.2in,height=2.6in]{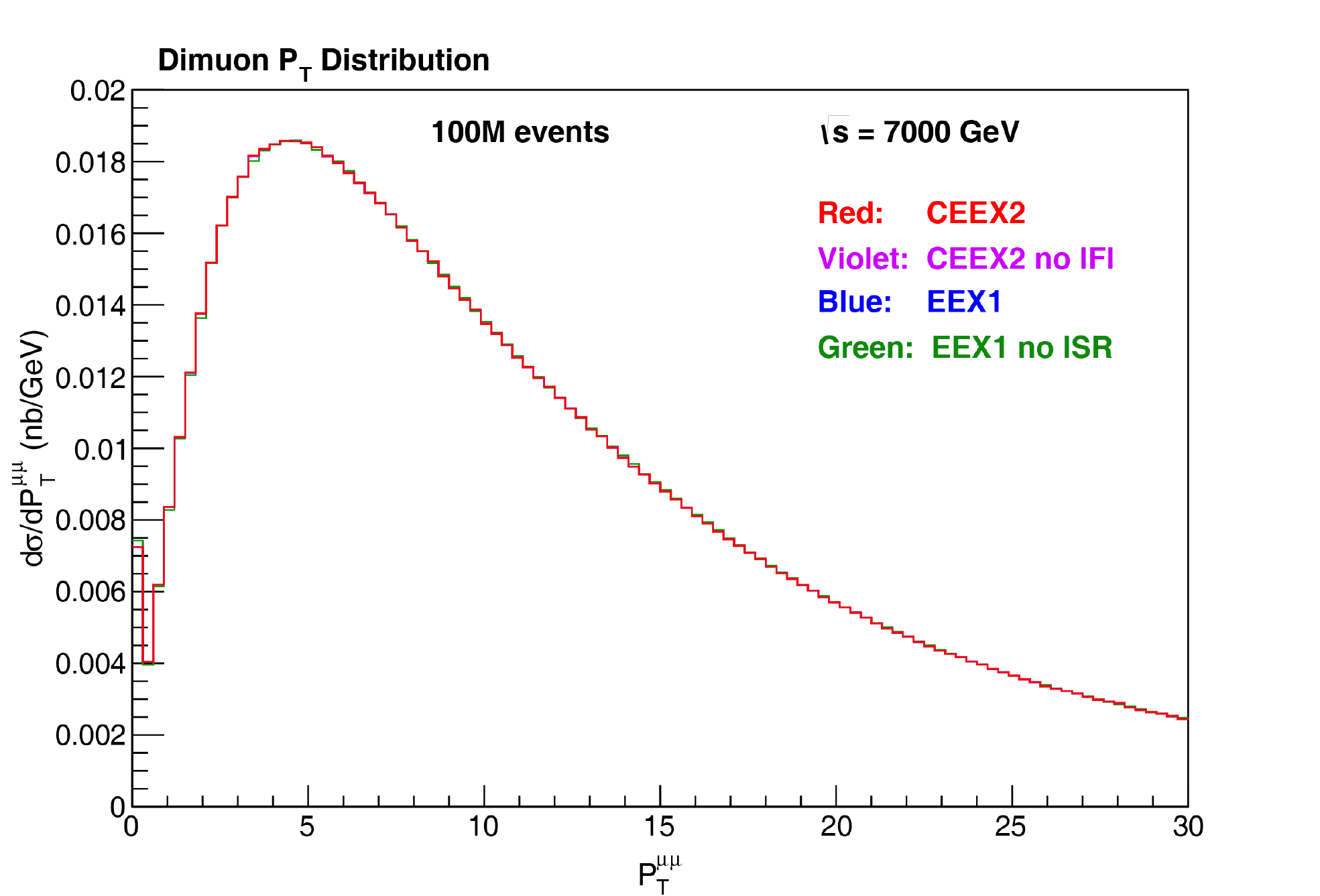}}
\put(3.2,0.5){\includegraphics[width=3.2in,height=2.6in]{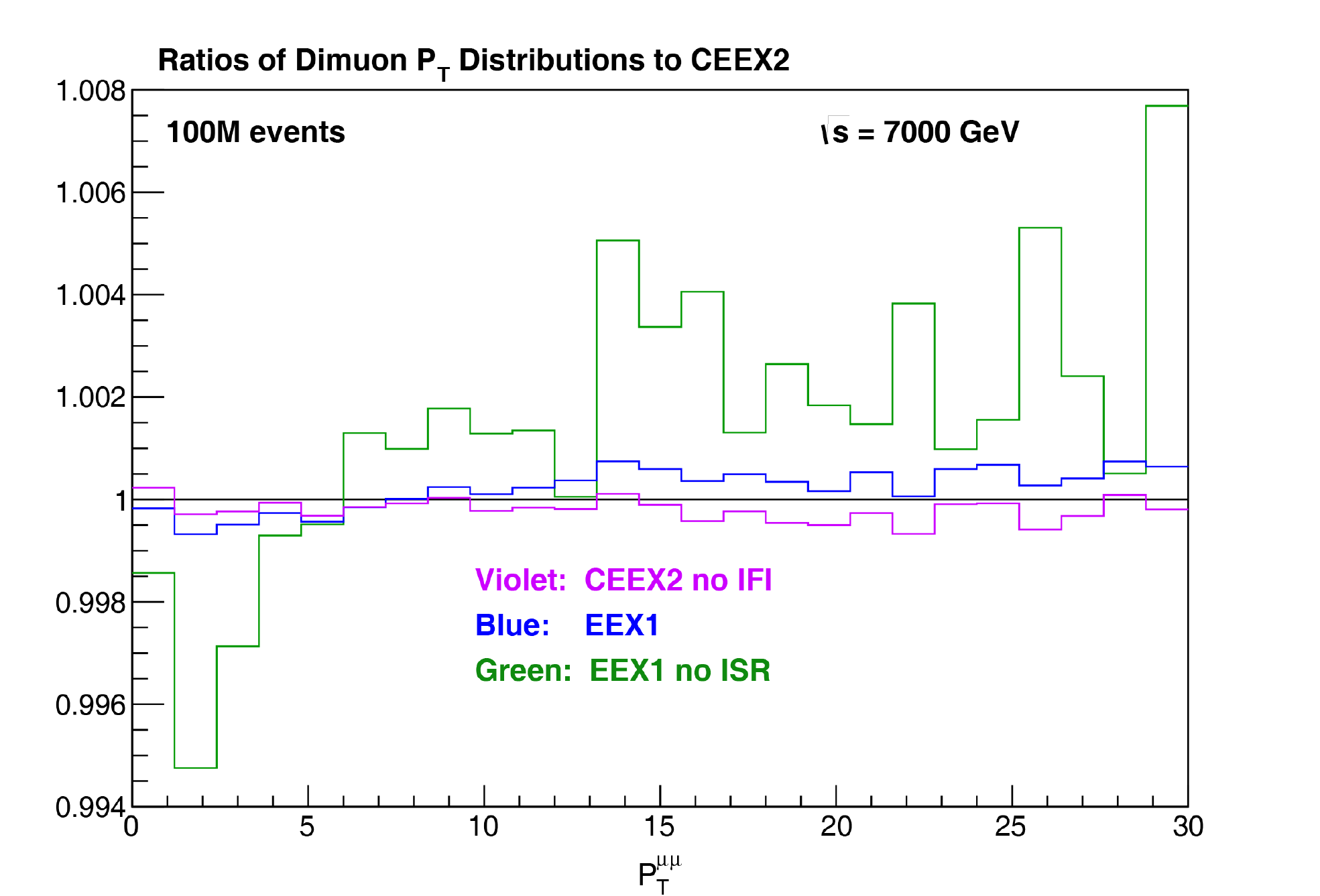}}
\end{picture}
\vspace{-0.75in}
\caption{Dimuon Transverse Momentum Distributions and Ratios}
\end{center}
\end{figure}

\begin{figure}[p]
\begin{center}
\setlength{\unitlength}{1in}
\begin{picture}(6.5,3.0)
\put(0,0.5){\includegraphics[width=3.2in,height=2.6in]{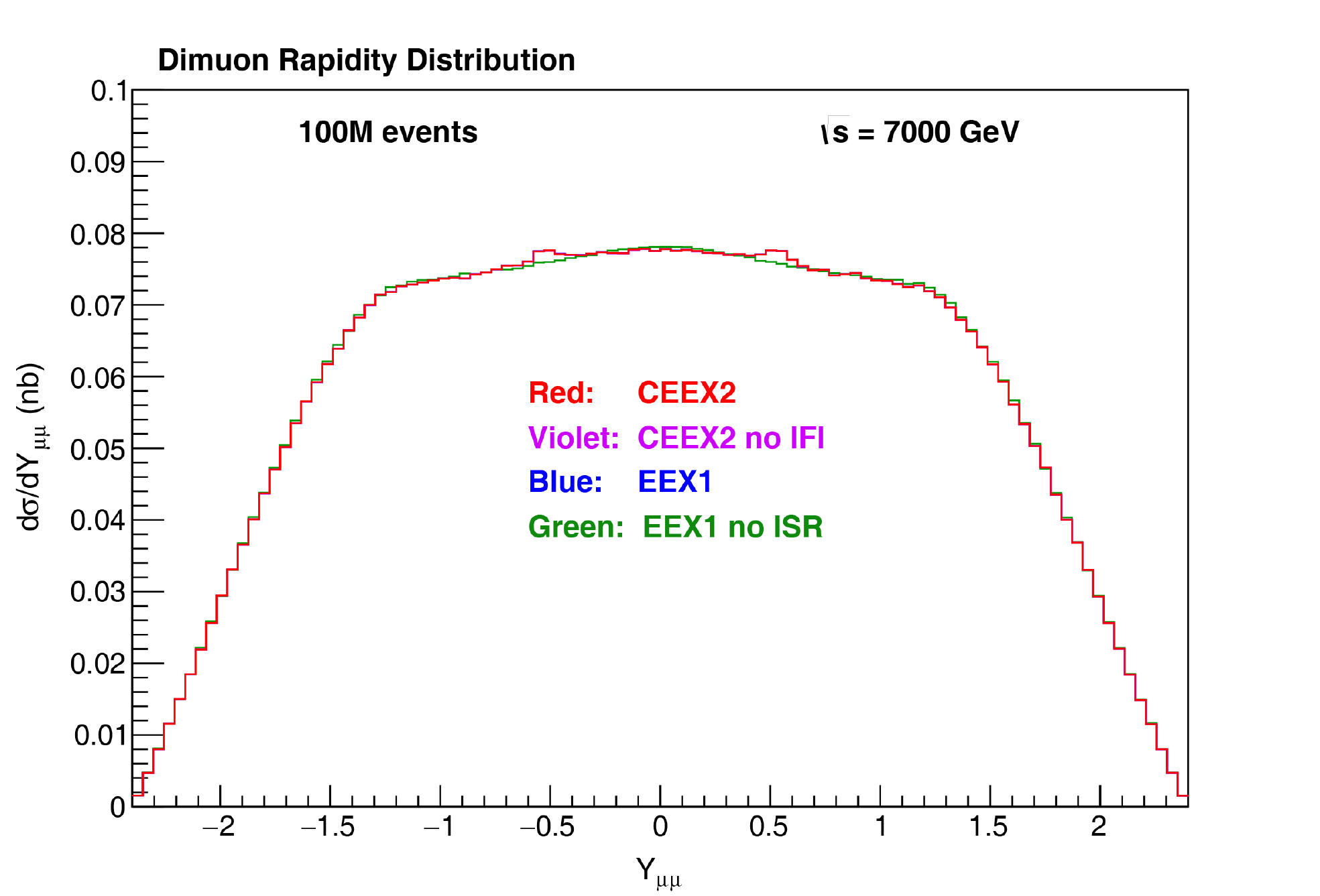}}
\put(3.2,0.5){\includegraphics[width=3.2in,height=2.6in]{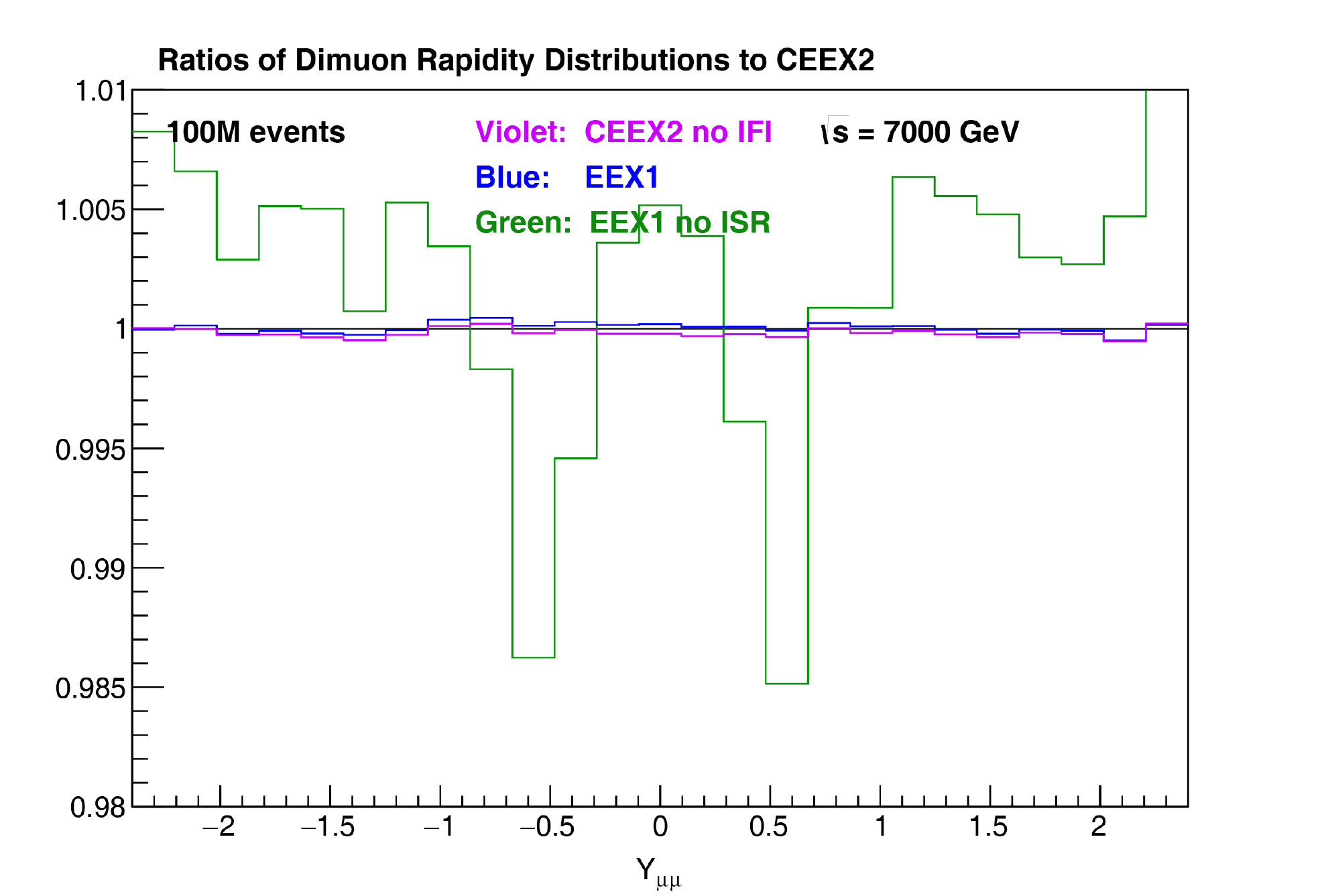}}
\end{picture}
\vspace{-0.75in}
\caption{Dimuon Rapidity Distributions and Ratios}
\end{center}
\end{figure}

\begin{figure}[ht!]
\begin{center}
\setlength{\unitlength}{1in}
\begin{picture}(6.5,3.0)
\put(0,0.5){\includegraphics[width=3.2in,height=2.6in]{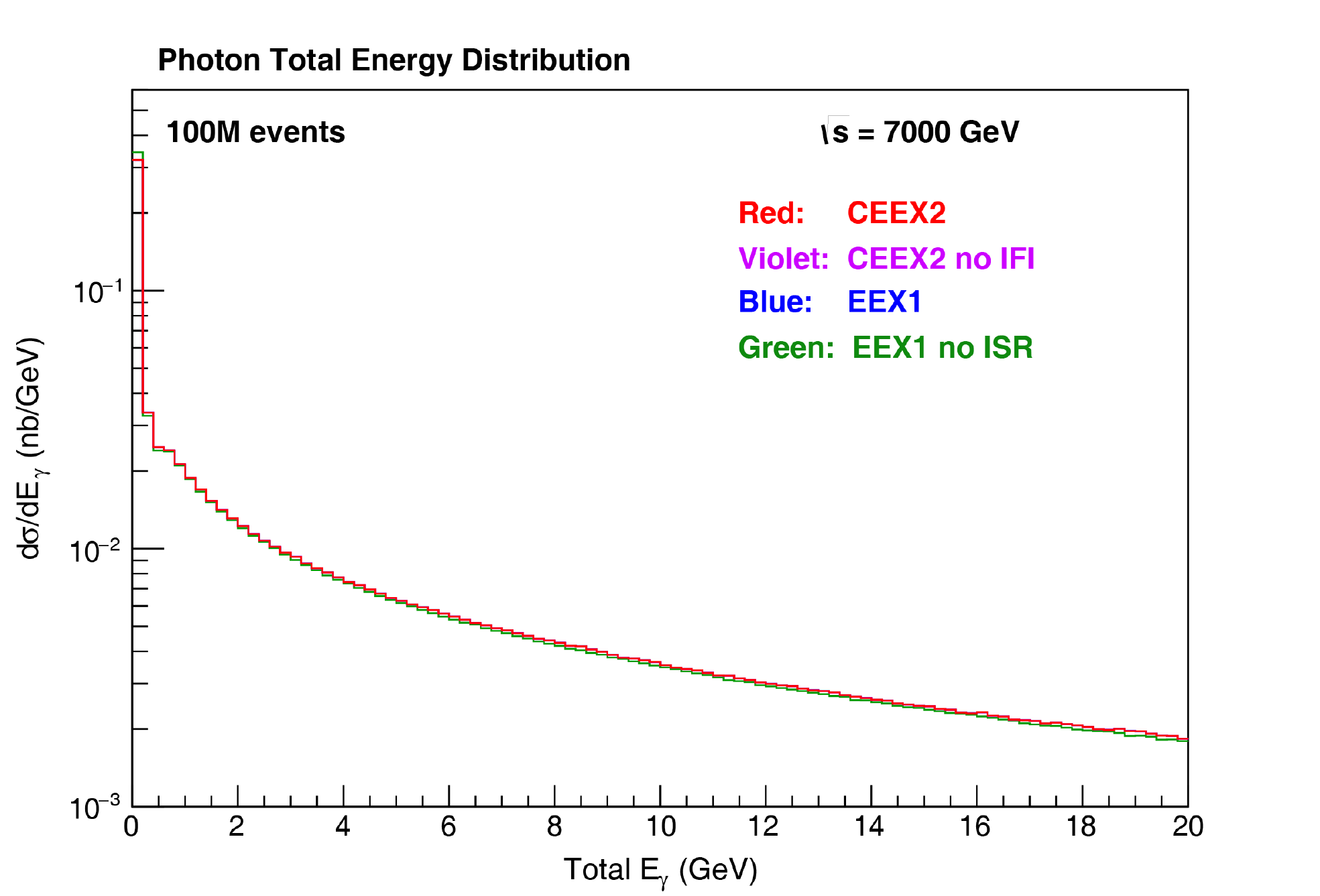}}
\put(3.2,0.5){\includegraphics[width=3.2in,height=2.6in]{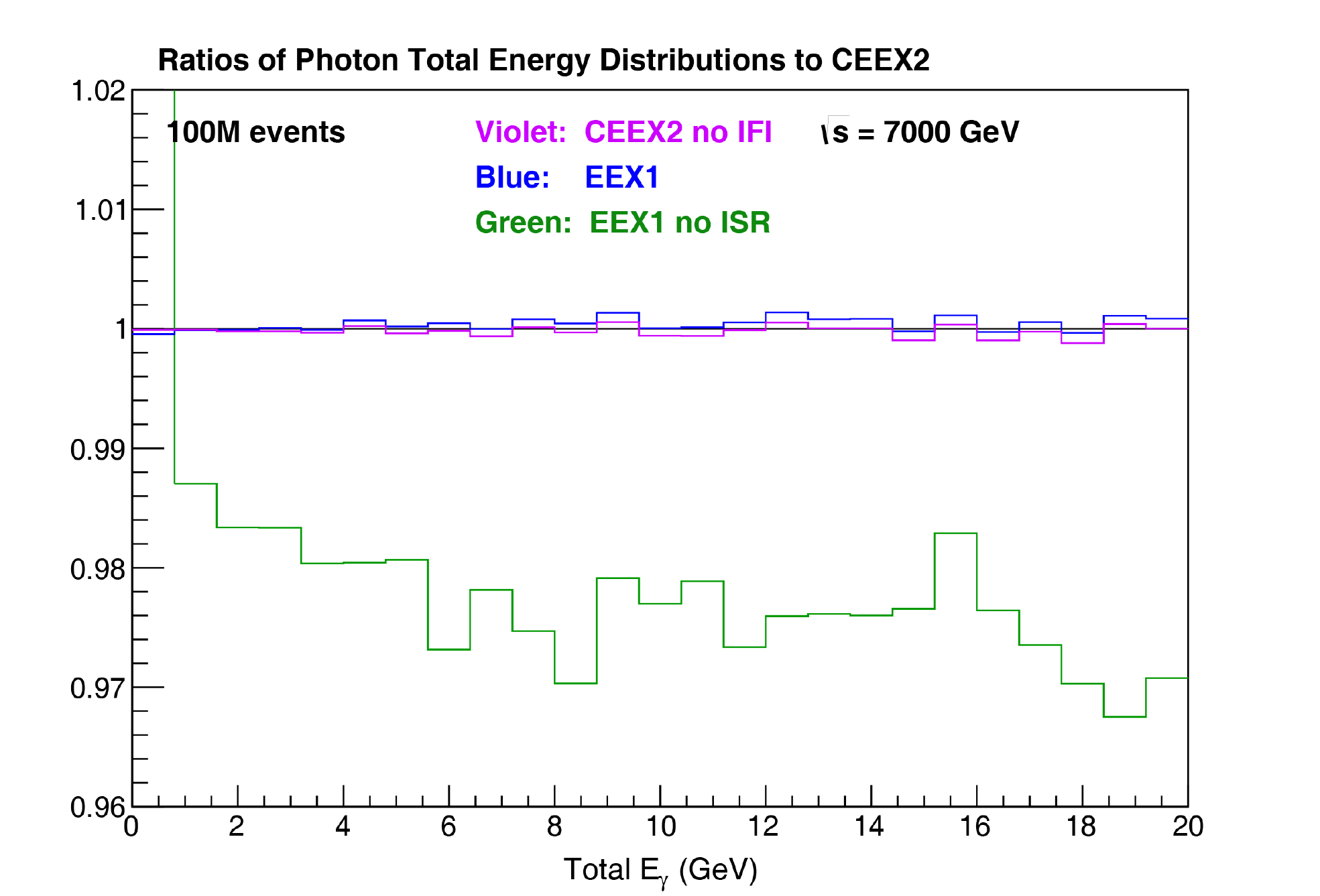}}
\end{picture}
\vspace{-0.75in}
\caption{Total Radiated Photon Energy}
\end{center}
\end{figure}

\begin{figure}[hb!]
\begin{center}
\setlength{\unitlength}{1in}
\begin{picture}(6.5,3.0)
\put(0,0.5){\includegraphics[width=3.2in,height=2.6in]{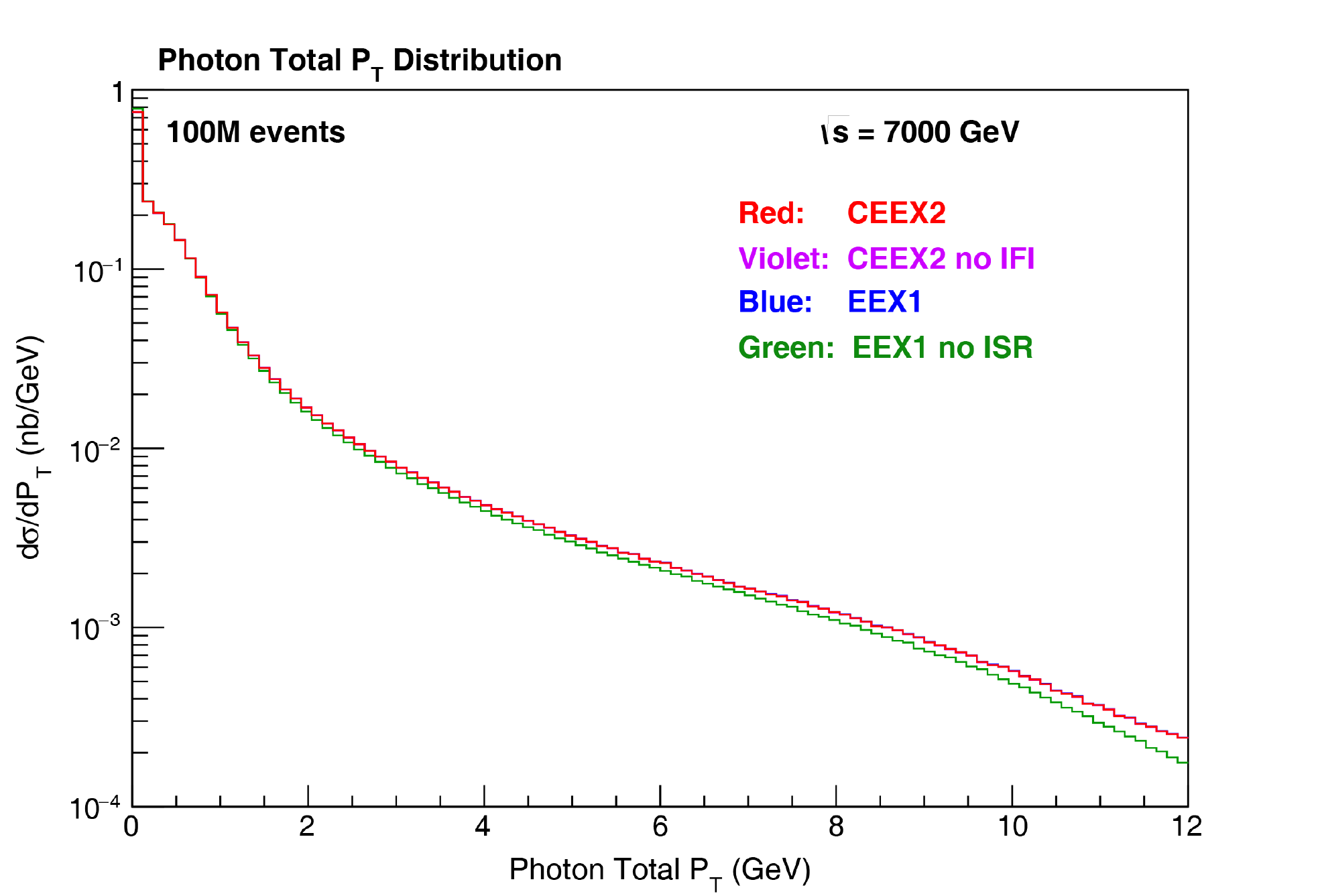}}
\put(3.2,0.5){\includegraphics[width=3.2in,height=2.6in]{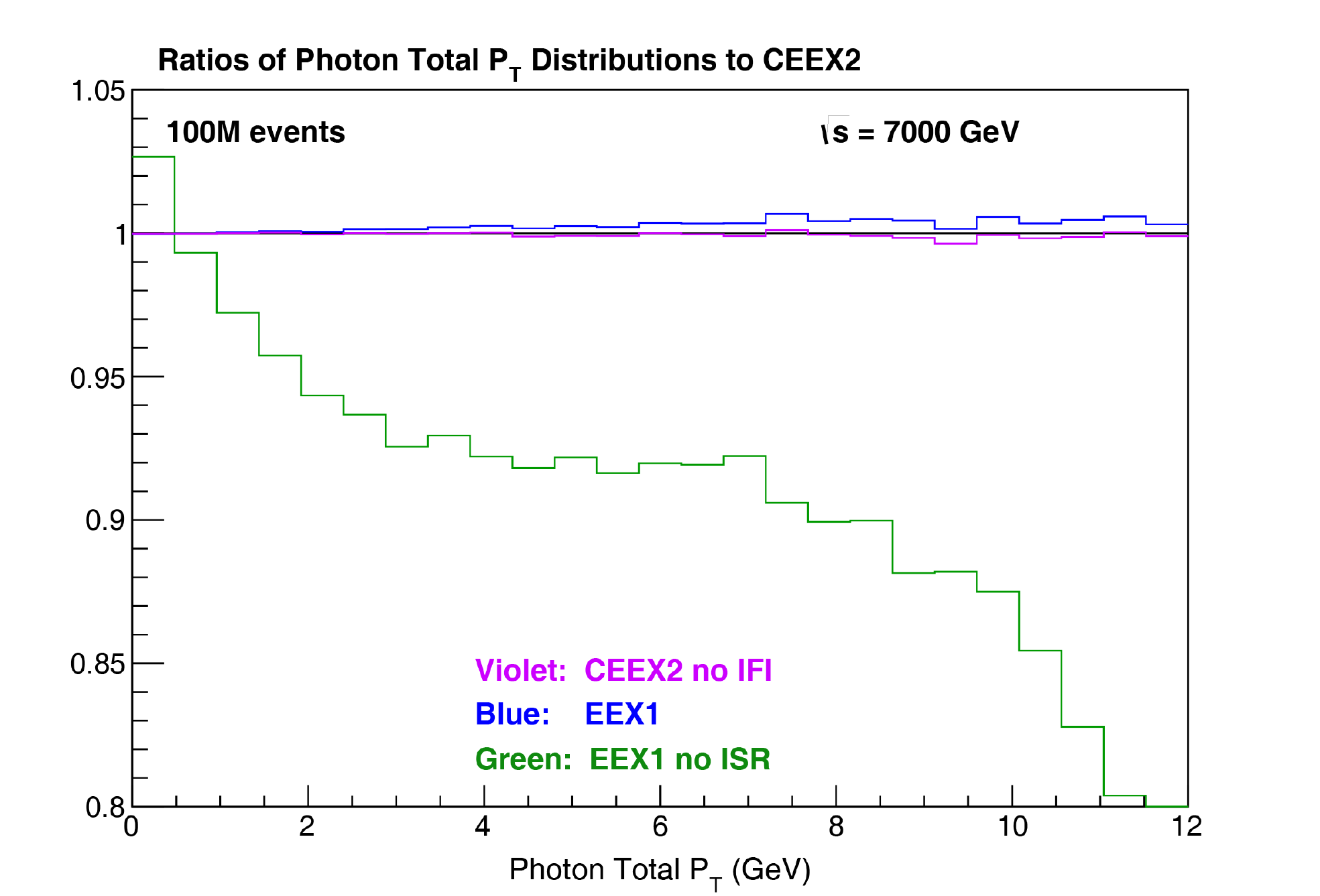}}
\end{picture}
\vspace{-0.75in}
\caption{Total Transverse Momentum of Radiated Photons}
\end{center}
\end{figure}
\clearpage
\section{Comparisons of \KK{MC}-hh and HORACE}
In this section, we show untuned comparisons of muon production at 8 TeV with 
MSTW PDFs and a cut 50 GeV $< M_{q\overline q} < $ 200 GeV on the generated 
quark pair and no QCD shower. We compare the \KK{MC}-hh results to
 HORACE,~\cite{Horace1,Horace2} which includes ${\cal O}(\alpha)$ EW 
corrections and exponentiated
final state photon emission.  The HORACE events are generated in the ``best''
EW scheme with exponentiated FSR. In this mode, HORACE should agree closely
with \KK{MC}-hh in its CEEX ${\cal O}(\alpha)$ exponentiated mode with ISR 
turned off.  Unshowered events from HERWIG6.5 are shown for a comparison without
EW corrections. The samples for HORACE and HERWIG6.5 have $10^8$ events, while
the \KK{MC}-hh sample has $25\times 10^6$ events. We include two \KK{MC}-hh
results:  
the full result including ISR, FSR, and IFI at ${\cal O}(\alpha^2 L)$ and a 
restriction to FSR only at ${\cal O}(\alpha)$ (CEEX1), which is should
be similar to HORACE. The HORACE Born-level result, without photons, is also
compared. Table 2 shows differences between the total cross sections
and the full \KK{MC}-hh result. All comparisons are without a QCD shower.

\vbox{
\begin{center}
\begin{tabular}{|l|c|c|c|}
\hline
MC	    	& EW Corrections 	& $\sigma$ (pb) & Difference\\
\hline
\KK{MC}-hh 	& CEEX2    	& $993\pm 1$   	& $\times$ \\
\KK{MC}-hh      & CEEX1 (no ISR)& $991\pm 1$ 	& $-0.20\%$\\
HORACE  	& ${\cal O}(\alpha)$ exp. & $1009.6\pm 0.4$ 	& $+1.7\%$\\
HORACE          & Born (no $\gamma$'s)    & $1025.2\pm 0.4$ 	& $+3.2\%$\\
HERWIG6.5       & Born (no $\gamma$'s)    & $1039.6\pm 0.2$     & $+4.7\%$\\
\hline
\end{tabular}
\\[1em]
{{\bf Table 2.} Total Cross Section Comparisons and Difference Relative to CEEX2}
\end{center}
}

Figure 8 compares the $\mu^-$ transverse momentum and pseudorapidity 
distributions.  Figure 9 compares the dimuon invariant mass distribution and 
the multiplicity of photons having at least 1 GeV of energy. 
There is a small normalization difference in the 
distributions, but they appear to be as compatible as can be expected prior to 
a precise tuning of the parameters. 

\begin{figure}[hb!]
\begin{center}
\setlength{\unitlength}{1in}
\begin{picture}(6.5,3.0)
\put(0,0.5){\includegraphics[width=3.2in,height=2.6in]{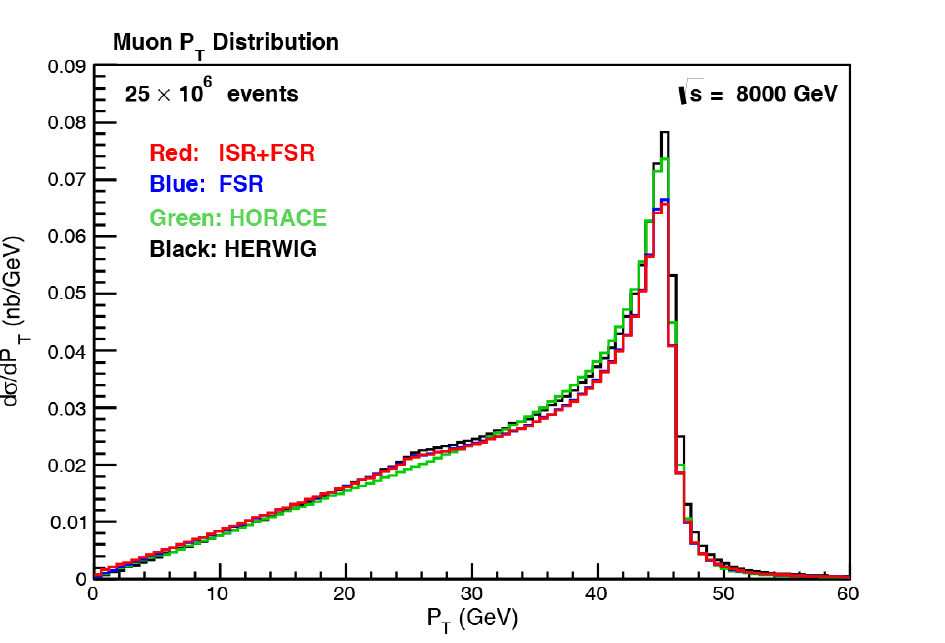}}
\put(3.2,0.5){\includegraphics[width=3.2in,height=2.6in]{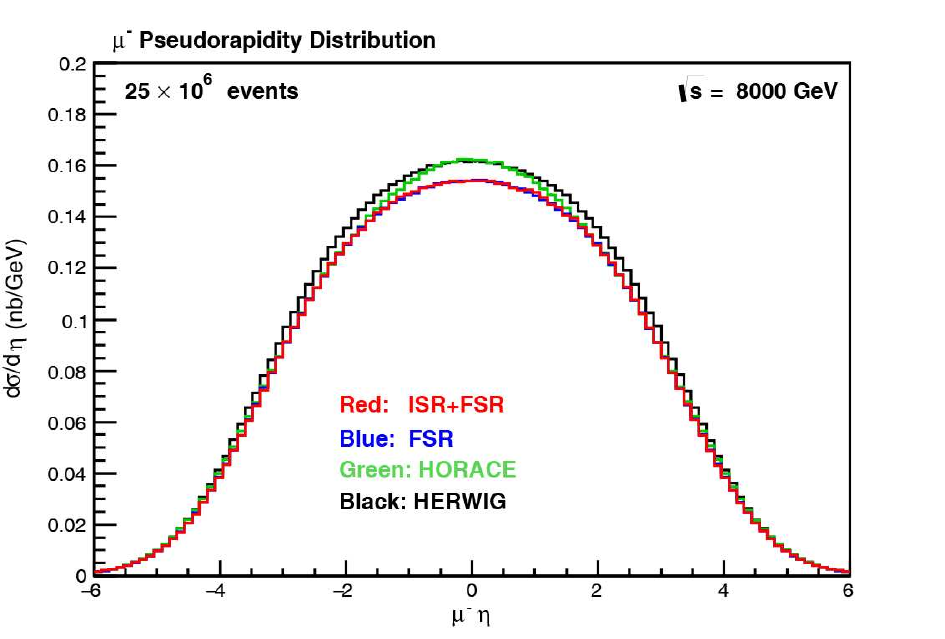}}
\end{picture}
\vspace{-0.75in}
\caption{Muon Transverse Momentum and Pseudorapidity Distributions }
\end{center}
\end{figure}

\begin{figure}[ht!]
\begin{center}
\setlength{\unitlength}{1in}
\begin{picture}(6.5,3.0)
\put(0,0.5){\includegraphics[width=3.2in,height=2.6in]{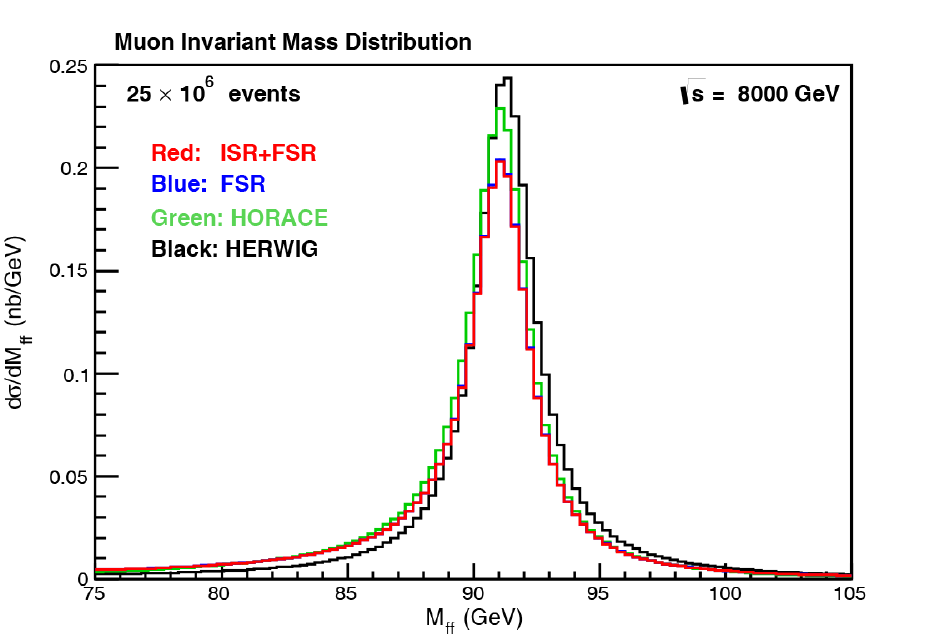}}
\put(3.2,0.5){\includegraphics[width=3.2in,height=2.6in]{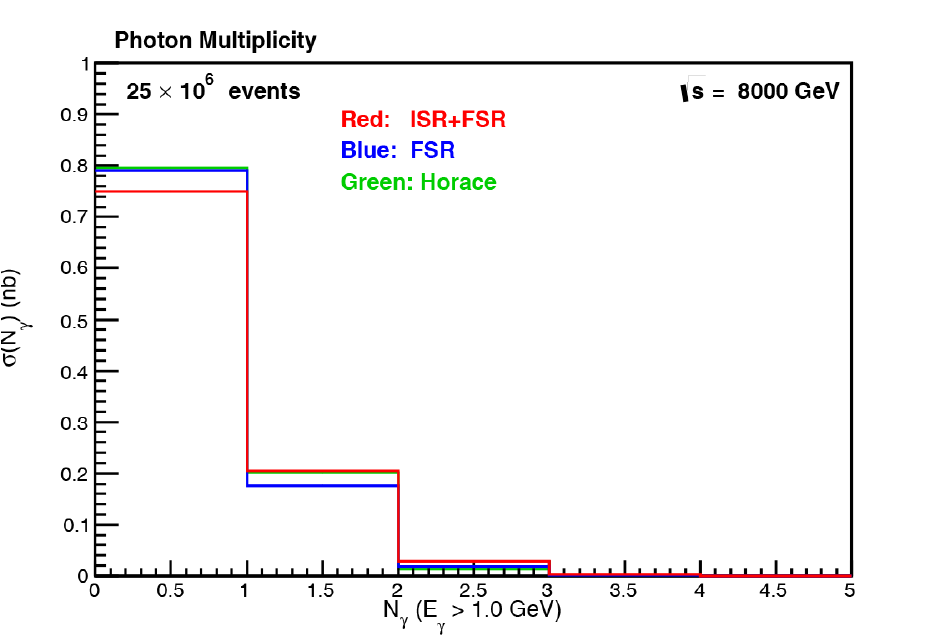}}
\end{picture}
\vspace{-0.75in}
\caption{Dimuon Invariant Mass and Photon Multiplicity (for $E_\gamma > 1$ GeV) Distributions}
\end{center}
\end{figure}


\section{Conclusions}

The results for ATLAS cuts show that for per mille level studies, ISR, IFI, and
exact ${\cal O}(\alpha^2 L)$ corrections should all be included for a 
conservative estimate of the precision tag. \KK{MC}-hh is available on request
for such studies.
Untuned comparisons to HORACE show a promising level of agreement when the
programs are compared at a compatible precision level (${\cal O}(\alpha)$ with
exponentiated FSR radiation). Tuned comparisons to the results in 
ref.\ \cite{vicini-wack} should be available in the near future. 

Further developments are under-way, including improvements to the interface
which will allow \KK{MC}-hh to operate either as a primary generator, 
creating events to be showered subsequently (the present mode), or to operate
as an add-on generator, to add photons and appropriate reweighting to 
events generated by any available parton shower, including NLO showers. It is
also anticipated that NLO QCD will be added directly to 
\KK{MC}-hh using the KrkNLO scheme.~\cite{krknlo} 
Other future enhancements may include 
adding the effect of additional fermion pairs and updating the
EW matrix element corrections using a version of 
SANC.~\cite{Andonov:2004hi,Andonov:2008ga} 
It should
be straightforward to create a version of \KK{MC}-hh for $W$ production, 
but this is not yet under development.

\end{document}